\documentclass[journal,transmag]{IEEEtran}

\usepackage{url}
\usepackage{amssymb}

\begin{document}

\IEEEtitleabstractindextext{%
\begin{abstract}
The desired result of magnetic anisotropy investigations is the
determination of value(s) of various anisotropy constant(s).  This is
sometimes difficult, especially when the precise knowledge of saturation
magnetization is required, as it happens in ferromagnetic resonance
(FMR) studies.  In such cases we usually resort to `trick' and fit our
experimental data to the quantity called \emph{anisotropy field}, which is
strictly proportional to the ratio of the searched anisotropy constant
and saturation magnetization.  Yet, this quantity is scalar,
simply a~number, and is therefore of little value for modeling
or simulations of the magnetostatic or micromagnetic structures.
Here we show how to `translate' the values of magnetic anisotropy
constants into the complete vector of  magnetic anisotropy field.
Our derivation is rigorous and covers the most often encountered cases,
from uniaxial to cubic anisotropy.
\end{abstract}
}

\title{ Where is magnetic anisotropy field pointing to?}%

\author{\IEEEauthorblockN{
Marek~W. Gutowski%\IEEEauthorrefmark{$\star$}
}%
\IEEEauthorblockA{%\IEEEauthorrefmark{$\star$}
Institute of Physics,
Polish Academy of Sciences, 02--668 Warszawa, Poland}
}%

%\thanks{Manuscript received December 25, 2013; revised December 25, 2013.
%Corresponding author: M.W.~Gutowskil (email: marek.gutowski@ifpan.edu.pl).}

%\markboth{Journal of \LaTeX\ Class Files,~Vol.~XX, No.~X, December~2013}%
%{Shell \MakeLowercase{\textit{et al.}}: Bare Demo of IEEEtran.cls for Journals}

%%\pacs{%Insert valid PACS here:
%%75.30.Gw, % Magnetic anisotropy
%%%%75.40.Mg, % Numerical simulation studies
%%75.78.Cd,  % Micromagnetic simulations
%%85.70.Ay   % Magnetic device characterization, design, and modeling
%%   }
% PACS, the Physics and Astronomy Classification Scheme.

\maketitle

\IEEEdisplaynontitleabstractindextext%

\IEEEpeerreviewmaketitle%

\section{Introduction}
\IEEEPARstart{I}{n technically} important simulations of magnetic systems
at meso- or macroscopic scale
one is confronted with the problem of finding the equilibrium orientation
of local magnetization in each part of a~system under study.\  Similarly to the
micromagnetic calculations, the searched orientation is the one that minimizes
the free energy of each small element of such a~system.
At mesoscopic scale this seems easy:
the free energy is nothing else but the potential energy of a magnetic moment
$\mathbf{M}$ in magnetic field $\mathbf{B}_{\mathrm{eff}}$, namely 
$F=-\mathbf{M}\cdot\mathbf{B}_{\mathrm{eff}}$.\  Consequently, the lowest energy
orientation of $\mathbf{M}$ is strictly parallel to $\mathbf{B}_{\mathrm{eff}}$.
In what follows, we will rather talk about the free energy density, and therefore
write $F=-\mathbf{M}\cdot\mathbf{B}_{\mathrm{eff}}$, where we retain the same
symbol $F$ for free energy density, but $\mathbf{M}$ denotes now the (local)
magnetization.

It has to be stressed that we have to speak about $\mathbf{B}_{\mathrm{eff}}$,
and not about $\mathbf{H}_{\mathrm{eff}}$, as of the effective field.\  This is in
contrast with micromagnetic calculations, where the distinction between $\mathbf{H}$
and $\mathbf{B}$ fields is less important.\
Inside the ferromagnetic bodies the fields $\mathbf{H}$ and $\mathbf{B}$ are not
proportional to each other, not even necessarily parallel, as they have to satisfy
the relation $\mathbf{B}=\mu_{0}\left(\mathbf{H}+\mathbf{M}\right)$, with
$\mathbf{M}\,\ne\,0$\ (in SI units).\  It is nothing unusual to observe that
the vectors $\mathbf{B}$ and $\mathbf{H}$ are very often nearly antiparallel.

\section{The uniaxial case}
We start with probably the most often encountered situation of a~single
easy axis magnetic anisotropy.\  While it is natural to think about some
crystal structures as being uniaxial, the same property is also given
to the well known Stoner-Wohlfarth particle.\ Even nominally amorphous
materials are sometimes successfully described as being magnetically
uniaxial, see for example \cite{Zuberek}.

The free energy density with uniaxial anisotropy is usually written as
\begin{equation}
F=K\sin^{2}\theta\quad\textrm{with}\quad K>0\label{first}
\end{equation}
or, equivalently (since $\sin^{2}\theta = 1 - \cos^{2}\theta$ and we are only
interested in orientation-dependent part), as
\begin{equation}
F=-K\cos^{2}\theta,\label{second}
\end{equation}
where $K$ (often written as $K_{\mathrm{u}}$) is the
uniaxial anisotropy constant, and the angle $\theta$ is made by vectors
$\mathbf{M}$ and easy axis direction $\mathbf{e}$ (unit vector).
We prefer the form (\ref{second}), since it exhibits similar angular
dependence as the potential magnetostatic energy usually written as
\begin{equation}
F=-\frac{1}{2}\,\mathbf{M}\cdot\mathbf{B} = -\frac{1}{2}\,MB\cos\theta,
\label{free}
\end{equation}
where the subscript "eff" has been dropped for sake of clarity.\ Besides, the
field $\mathbf{B}$ above is the one generated by all surrounding magnetic
moments, not by the object under study.\  This is why the factor $1/2$ appears
in formula (\ref{free}).\  Neglect of self-interaction is justified, as it
by no means depends on orientation of the object.\ In this
spirit we may rewrite the Eq.~\ref{second}:
\begin{equation}
F=-K\left(\frac{\mathbf{M}\cdot\mathbf{e}}
               {\left|\mathbf{M}\right|\cdot\left|\mathbf{e}\right|}\right)^{2}
 =-\frac{K}{M^{2}} \left(\mathbf{M}\cdot\mathbf{e}\right)^{2}
\label{sqform}
\end{equation}
Equating (\ref{free}) and (\ref{sqform}) we obtain
\begin{equation}
-\frac{1}{2}\,\mathbf{M}\cdot\mathbf{B}=-\frac{K}{M^{2}}
 \left(\mathbf{M}\cdot\mathbf{e}\right) \left(\mathbf{M}\cdot\mathbf{e}\right)
\end{equation}
and thus immediately
\begin{equation}
\mathbf{B} = \frac{2K}{M} \left(
\frac{\mathbf{M}}{M}\cdot\mathbf{e}\right)\,\mathbf{e}=
\frac{2K}{M}\cos\theta\cdot{\mathbf e}
\end{equation}
As expected, $\mathbf{B}\parallel\mathbf{e}$\ and\ $B\,\propto\,\frac{2K}{M}$.~\
Moreover, $\mathbf{B}$ is insensitive to easy axis reversal/reflection\
$\mathbf{e}\rightarrow\mathbf{-e}$ --- as it should be.\
This result is in full agreement with \emph{customary
definition} \cite{RS} of anisotropy field as $H_{a} = \frac{2K}{\mu_{0}M}$.\
On the other hand one can see that the magnitude of anisotropy field
depends on orientation of local magnetization with respect to local easy axis,
ranging from $B=0$ for $\mathbf{M}\perp\mathbf{e}$ to $B=\frac{2K}{M}$ when
$\mathbf{M}\parallel\mathbf{e}$.

In what follows we will be using shortened notation: $\mathbf{M}/M=\mathbf{m}$,
and also $\mathbf{m}\cdot\mathbf{e}=\cos\theta$.

\section{More than one easy axis}
The FMR spectral features in presence of exactly two different (i.e. non-parallel)
easy axes were extensively investigated by Cochran and Kambersky \cite{Kambersky}.\
They were interested in surface anisotropy of ultrathin layers, which may
differ for substrate-side and free side of a~layer.

Our extension of the earlier described procedure for samples having even more
easy axes \{$\mathbf{e}_{1}, \mathbf{e}_{2}, \ldots\,$\} is straightforward,
if only their anisotropy may be written as a~sum
\begin{equation}
 F =  K^{(1)}\sin^{2}\theta_{1} + K^{(2)}\sin^{2}\theta_{2} + \cdots\\
\end{equation}
or, equivalently
\begin{equation}
 F =  -K^{(1)}\cos^{2}\theta_{1} - K^{(2)}\cos^{2}\theta_{2} - \cdots
\end{equation}
We have to stress that all anisotropy constants, $K^{(i)}$, have to be positive,
otherwise we deal with easy plane(s) rather than with easy magnetization
directions.\ Anisotropy field is then
\begin{equation}
\mathbf{B}=\frac{2K^{(1)}}{M}\left(
\mathbf{m}\cdot\mathbf{e}_{1}\right)\mathbf{e}_{1} +
             \frac{2K^{(2)}}{M}\left(
             \mathbf{m}\cdot\mathbf{e}_{2}\right)\mathbf{e}_{2} + \cdots
\end{equation}
The last formula applies to thin films of cubic, tetragonal or hexagonal symmetry,
as well as to strained amorphous materials, whenever higher order anisotropy
terms (i.e. containing terms like $K^{(4)}\sin^{4}\theta$) can be neglected.\ It is
worth noticing that presence of more than one easy axes makes orientation
of the anisotropy field a~linear combination of easy directions.\
Thus generally $\mathbf{B}\!\nparallel\!\mathbf{e_{1}}$,
$\mathbf{B}\!\nparallel\!\mathbf{e_{2}}$, and so on, even if\  $K^{(1)}=K^{(2)}$.\
A~very special case will occur when easy axes are perpendicular to each
other, see later discussion concerning cubic anisotropy case.

\section{Higher order single easy axis}
The magnetocrystalline anisotropy of bulk tetragonal and hexagonal materials
is often written as~\cite{wiki}:
\begin{equation}
F=K_{1}\sin^{2}\theta + K_{2}\sin^{4}\theta.
\label{higher-order}
\end{equation}
Proceeding as before we obtain
\begin{equation}
F=\left[-\left(K_{1}+2K_{2}\right)\cos\theta + K_{2}\cos^{3}\theta\right]
\left(\mathbf{m}\cdot\mathbf{e}\right)/M
\end{equation}
and anisotropy field takes the form
\begin{eqnarray}
\mathbf{B}&\!=\!&\frac{2}{M}\left[\left(K_{1}+2K_{2}\right)\cos\theta
-K_{2}\cos^{3}\theta \right]\mathbf{e}\nonumber\\
&\!=\!&\frac{2}{M}\left[\left(K_{1}+2K_{2}\right)\left(\mathbf{m}\cdot\mathbf{e}\right)
-K_{2}\left(\mathbf{m}\cdot\mathbf{e}\right)^{3} \right]\mathbf{e}
\label{hord}
\end{eqnarray}
Again, like in single easy axis case, $\mathbf{B}\!\parallel\!\mathbf{e}$, with magnitude
varying with orientation of magnetization $\mathbf{M}$.

\subsection{Non-positive anisotropy constants}
Formula (\ref{hord}) was derived somewhat mechanically, without paying much
attention to the magnitudes and signs of anisotropy constants $K_{1}$ and
$K_{2}$.\ In order to find the possible equilibrium positions of the vector
$\mathbf{m}$ one has to solve the equation $\partial{F}/\partial{\theta}=0$
for $\theta\in[0,\pi]$.\   It is easy to see that
there may be up to $4$ possible solutions (not necessarily corresponding
to the local free energy minima!)
\begin{equation}
\theta_{1}=0,~\
\theta_{2}=\frac{\pi}{2},~\
\theta_{3}=\pi,~\ \textrm{or}~\
\sin^{2}\theta_{4}=-\frac{K_{1}}{2K_{2}}\label{solutions}
\end{equation}
The first three solutions always exist, while the fourth is~obviously only
possible when
\begin{equation}
0<\,-K_{1}/\left(2K_{2}\right)\,<\,1\label{conecond}
\end{equation}
(we write sharp inequalities,
since the limiting cases coincide with first three solutions).\  The fourth
solution is interesting in itself as it corresponds to the not so frequent
case of easy cone magnetization.\

But which solution corresponds to our particular case?  Of course, the one
making the global minimum of the free energy density $F$.\  The values
of $F$ corresponding to each solution are:
\begin{eqnarray}
F\left(\theta=0\right)&=&F\left(\theta=\pi\right) = 0\label{e-axis}\\
F\left(\theta=\frac{\pi}{2}\right)&=&K_{1}+K_{2}\label{e-plane}\\
F\left(\theta_{4}\right)&=&-\frac{K_{1}^{2}}{4K_{2}}\label{e-cone}
\end{eqnarray}

The necessary condition for (local!) minimum of $F$ is:
$\partial^{2}{F}/\partial\theta^{2}\,>\,0$.\   Thus for easy axis case
($\theta=0$ or $\theta=\pi$) we should have
\begin{equation}
K_{1}>0,
\end{equation}
for easy plane ($\theta=\pi/2$):
\begin{equation}
-K_{1}-2K_{2} > 0,
\end{equation}
while for easy cone
\begin{equation}
-4K_{1}\left(\frac{K_{1}}{2K_{2}}+1\right) > 0.\label{conefirst}
\end{equation}
Looking at inequality (\ref{conecond}) we immediately
conclude that expression contained in parentheses in (\ref{conefirst})
must be positive, thus the easy cone may only be realized when
\begin{equation}
K_{1}<0
\end{equation}
We stop the discussion concerning the correspondence between
values and signs of anisotropy constants $K_{1}$ and $K_{2}$ and the kind
of~magnetization of a~sample.\ Here we are interested only in the
direction and magnitude of an~anisotropy field. 

\subsection{Conclusion} 
The formula (\ref{hord})
produces the correct result in every case, regardless of the relations
between the anisotropy constants $K_{1}$ and $K_{2}$.\  A~surprising part
of our result is the fact that rotational symmetry axis $\mathbf{e}$
(of $C_{\infty}$ type) of a~system dictates the alignment of the anisotropy
field, equally well for easy-axis, easy-plane or easy-cone magnetization.

To decide which type of equilibrium position (easy axis, easy plane or easy cone)
is currently at play, one has to carefully analyze the conditions from (\ref{solutions})
to (\ref{conefirst}).\  Of course, we are always looking for the global minimum of
a~free energy density, $F$.

\section{Curved surface case}
The case of otherwise amorphous ferromagnetic microwire is specific.\  In addition
to the energy given in Eq.~(\ref{higher-order}) we should consider one more, rather
unusual term\cite{MWG}, namely
\begin{equation}
F_{s}=K_{s}\,\left|\cos\theta\right|
\label{surfani}
\end{equation}
It is specific as the constant $K_{s}$  is always positive and inversely proportional
to the wire's diameter squared.\ For this reason it promotes an easy plane,
perpendicular to the wire, rather than easy axis.\  Therefore, using its hard axis,
i.e. the one parallel to wire length, $\mathbf{e}$, the anisotropy field
in Eq.~(\ref{hord}) should be appended with
\begin{equation}
\mathbf{B}=-\frac{2K_{s}}{M}\,\mathrm{sign}
\left(\mathbf{m}\cdot\mathbf{e}\right)\,\mathbf{e}
\end{equation}
Here we have exploited the identity $\left|x\right|=x\cdot\mathrm{sign}(x)$.
Note the minus sign and that again $\mathbf{B}\!\parallel\!\mathbf{e}$.

It is interesting that the surface-generated anisotropy described by formula
(\ref{surfani}) applies equally well to rippled MBE grown surfaces~\cite{JNT},
not just to microwires only.

\section{Cubic anisotropy}
In cubic materials the density of anisotropic part of the free energy
takes the shape:
\begin{eqnarray}
F&=&\phantom{+}\,K_{1}\left(\alpha^{2}\beta^{2}
                                        +\beta^{2}\gamma^{2}
                                        +\gamma^{2}\alpha^{2}\right)\nonumber\\
 &&+\, K_{2}\,\alpha^{2}\beta^{2}\gamma^{2}\nonumber\\
 &&+\, K_{3}\left(\alpha^{4}\beta^{4}
                     + \beta^{4}\gamma^{4}
                     + \gamma^{4}\alpha^{4} \right) + \cdots
\label{cubic}
\end{eqnarray}
Here $\alpha, \beta$, and $\gamma$ are direction cosines
in Cartesian coordinate frame.\ Depending on relation between $K_{1}$
and $K_{2}$, the easy axes are $\left[100\right]$ or $\left[111\right]$\ when
$K_{1}>0$,\  while for $K_{1}<0$ the easy axes are aligned
with $\left[111\right]$ or $\left[110\right]$ directions~\cite{wiki}.\
Under no circumstances easy cone or easy plane appears.\  
This discussion, strictly speaking, is only valid when $K_{3}$, and higher
terms of magnetic anisotropy are negligible.

From now on we will not pay attention to the relations between $K_{1},\,K_{2},\,$
and $K_{3}$.  Our goal is to find the orientation and strength of local
magnetocrystalline anisotropy field.\  The natural choice for three orthogonal
directions is to take as `easy' axes the ones coinciding with the coordinate
frame, namely:
$\mathbf{e_{1}}=\left(100\right)$, $\mathbf{e_{2}}=\left(010\right)$,
and $\mathbf{e_{3}}=\left(001\right)$.\  Needless to say that other three
orthogonal, highly symmetrical,  directions may be used as well.\
For this reason, in the following, we will use only their general symbols,
i.e. $\mathbf{e_{1}},\,\mathbf{e_{2}},\,$ and $\mathbf{e_{3}}$.

We will analyze the formula (\ref{cubic}) term after term, $F=F_{1}+F_{2}+F_{3}$,
remembering that
$\alpha=\mathbf{m}\cdot\mathbf{e}_{1}$, $\beta=\mathbf{m}\cdot\mathbf{e}_{2}$,
and $\gamma=\mathbf{m}\cdot\mathbf{e}_{3}$.\

Writing the first term (of fourth order) twice and rearranging
the new expression we obtain
$$
F_{1}=\frac{K_{1}}{2}\left(\alpha^{2}\beta^{2}+
\underline{\beta^{2}\gamma^{2}}+
\underbrace{\gamma^{2}\alpha^{2}}
+\underline{\alpha^{2}\beta^{2}}+
\underbrace{\beta^{2}\gamma^{2}}
+\gamma^{2}\alpha^{2}\right)
$$
%\end{equation}
\begin{equation}
=\frac{K_{1}}{2}\left[
\alpha^{2}\left(\beta^{2}+\gamma^{2}\right)
+\beta^{2}\left(\alpha^{2}+\gamma^{2}\right)
+\gamma^{2}\left(\alpha^{2}+\beta^{2}\right)\right],
\end{equation}
and consequently{\small
\begin{eqnarray}
\mathbf{B}^{(1)}&\!=\!&-\frac{K_{1}}{M}\left\lbrace\left[\left(\mathbf{m}\cdot\mathbf{e}_{2}
\right)^{2}
+ \left(\mathbf{m}\cdot\mathbf{e}_{3}\right)^{2}\right]\left(\mathbf{m}\cdot\mathbf{e}_{1}
\right)\,
\mathbf{e}_{1}\right.\nonumber\\
&&\,\qquad+\left[\left(\mathbf{m}\cdot\mathbf{e}_{1}\right)^{2}
+ \left(\mathbf{m}\cdot\mathbf{e}_{3}\right)^{2}\right]\left(\mathbf{m}\cdot\mathbf{e}_{2}
\right)\,
\mathbf{e}_{2} \nonumber\\
&&\qquad+\left.\left[\left(\mathbf{m}\cdot\mathbf{e}_{1}\right)^{2}
+ \left(\mathbf{m}\cdot\mathbf{e}_{2}\right)^{2}\right]\left(\mathbf{m}\cdot\mathbf{e}_{3}
\right)\,
\mathbf{e}_{3}\right\rbrace%\nonumber
\end{eqnarray}}

The second term of (\ref{cubic}) needs triplication in order to extract
a~part linear in $\mathbf{m}$ from free energy density:
\begin{equation}
F_{2}=\frac{K_{2}}{3}\,\left(\alpha^{2}\beta^{2}\gamma^{2}
+\alpha^{2}\beta^{2}\gamma^{2}
+\alpha^{2}\beta^{2}\gamma^{2}\right)
\end{equation}
The corresponding anisotropy field is then:{\small
\begin{eqnarray}
\mathbf{B}^{(2)}&\!=\!&-\frac{2K_{2}}{3M}\left\lbrace\left[\left(\mathbf{m}\cdot\mathbf{e}_{2}\right)
\left(\mathbf{m}\cdot\mathbf{e}_{3}\right)\right]^{2}\left(\mathbf{m}\cdot\mathbf{e}_{1}\right)\,
\mathbf{e}_{1}\right.\nonumber\\
&&\qquad\phantom{(}+\left[\left(\mathbf{m}\cdot\mathbf{e}_{1}\right)
\left(\mathbf{m}\cdot\mathbf{e}_{3}\right)\right]^{2}\left(\mathbf{m}\cdot\mathbf{e}_{2}\right)\,
\mathbf{e}_{2} \nonumber\\
&&\!\qquad\phantom{(}+\left.\left[\left(\mathbf{m}\cdot\mathbf{e}_{1}\right)
\left(\mathbf{m}\cdot\mathbf{e}_{2}\right)\right]^{2}\left(\mathbf{m}\cdot\mathbf{e}_{3}\right)\,
\mathbf{e}_{3} \right\rbrace%\nonumber
\end{eqnarray}}
The eight order term again should be virtually doubled and rearranged,
leading finally to:{\small
\begin{eqnarray}
\mathbf{B}^{(3)}&\!=\!&-\frac{K_{3}}{M}\left\lbrace\left[\left(\mathbf{m}\cdot\mathbf{e}_{2}
\right)^{4}
+ \left(\mathbf{m}\cdot\mathbf{e}_{3}\right)^{4}\right]\left(\mathbf{m}\cdot\mathbf{e}_{1}
\right)^{3}\,
\mathbf{e}_{1}\right.\nonumber\\
&&\,\qquad+\left[\left(\mathbf{m}\cdot\mathbf{e}_{1}\right)^{4}
+ \left(\mathbf{m}\cdot\mathbf{e}_{3}\right)^{4}\right]\left(\mathbf{m}\cdot\mathbf{e}_{2}
\right)^{3}\,
\mathbf{e}_{2} \nonumber\\
&&\qquad+\left.\left[\left(\mathbf{m}\cdot\mathbf{e}_{1}\right)^{4}
+ \left(\mathbf{m}\cdot\mathbf{e}_{2}\right)^{4}\right]\left(\mathbf{m}\cdot\mathbf{e}_{3}
\right)^{3}\,
\mathbf{e}_{3}\right\rbrace%\nonumber
\end{eqnarray}}
The full anisotropy field is, of course, equal to the sum $\mathbf{B}=
\mathbf{B}^{(1)}+ \mathbf{B}^{(2)} + \mathbf{B}^{(3)}$.
The dependence of $\mathbf{B}$ on $\mathbf{m}$ appears highly non-linear
and pretty complex.\ Nevertheless  $\mathbf{B}(\mathbf{-m})=
-\mathbf{B}(\mathbf{m})$, as expected.\ 

It is interesting to see what happens when the local magnetization vector,
$\mathbf{m}$, is oriented along one of the directions $\mathbf{e}_{i}$, $i=1,2,3$.\
Then $\left|\mathbf{m}\cdot\mathbf{e}_{i}\right|=1$ but for any $j\ne{i}$\ we
necessarily  have
$\mathbf{m}\cdot\mathbf{e}_{j}=0$, hence
$\mathbf{B}=\mathbf{B}^{(1)} + \mathbf{B}^{(2)} + \mathbf{B}^{(3)}=0$ -- regardless
of the values of $K_{1}, K_{2},$ and $K_{3}$.\  High symmetry cubic directions
are always special: they point either to extrema or to saddles of free energy
density.

\section{Ending remarks}
We have shown how to convert the various phenomenological expressions
for magnetocrystalline free energy density into a~vector of the so called
\emph{anisotropy field}.\ This fictitious field has nothing to do with external
field, nor with dipole-type field generated by the other parts of a~sample.\
It is nevertheless very useful during simulations based on Landau-Lifshitz-Gilbert
(LLG) equation of motion~\cite{OOMMF,MUMAX,VAMP}, or some variants
of Monte Carlo approaches as well.

\ifCLASSOPTIONcaptionsoff%
  \newpage
\fi

\end{document}